\documentclass[12pt,a4]{article}
\usepackage{psfig,epsfig,graphicx}
\usepackage{amsmath}
\usepackage{amssymb}
\usepackage{cite}

\begin{document}

\title{Shear viscosity and entropy production related to viscous process in hot QGP at finite density}
\author{Liu Hui \thanks{liuhui@iopp.ccnu.edu.cn} \ \ Hou Defu \thanks{hdf@iopp.ccnu.edu.cn}
 \ \ Li Jiarong \thanks{ljr@iopp.ccnu.edu.cn} \\[0.5cm] {\it\small Institute of Particle Physics, Central China Normal University,}\\ {\it
\small Wuhan(430079), P.R.China}}
\date{}
\maketitle
\begin{abstract}
In the framework of irreversible thermodynamics, we studied the
transport properties of QGP. Shear viscosity and non-equilibrium
entropy density related to viscous process at finite density has
been investigated in weakly coupled limit by using kinetic theory.
The results show that the chemical potential increases viscosity
yet decreases the non-equilibrium entropy density and thus
contributes positively to their ratio compared to the pure
temperature case. As to the temperature dependence of the ratio,
the ratio first decreases rapidly and then increases in the
physical region, presenting a minimum value of 0.4 at the
temperature around 182MeV.
\end{abstract}

\section{Introduction}

Recently scientists believe that the quark-gluon plasma(QGP) found
at relativistic heavy ion collider(RHIC) is strongly coupled,
contrasting with the weakly coupled as expected before, which is
the so-called sQGP\cite{Shuryak,Gyulassy,Muller}. They also
believe when the temperature goes higher, for example $T>2-4 T_c$,
the hot medium would be expected to become a weakly coupled system
with some dissipative structures. Additionally, investigations on
compact star demonstrate that the viscous properties are
significant in explaining many of their
behaviors\cite{Haensel,Xiaoping}. Therefore dissipative structure
of strong interaction system, especially the sQGP, is a remarkable
topic.

In irreversible thermodynamics, the discussions on the dissipative
properties of a system are focused on the entropy production in an
unit time
\begin{equation}\label{delta s}
\triangle S=\sum_i T_i X_i
\end{equation}
where $X_i$ is the thermal force which is determined by the
gradients of energy, temperature, chemical potential {\it etc}.
$T_i$ is the corresponding flow driven by $X_i$ which can be
written in the linear response approximation as
\begin{equation}\label{flow expansion}
T_i=\sum_j L_{ij}X_j
\end{equation}
where $L_{ij}$ are the transport coefficients. Thus it can be seen
clearly that the entropy production is determined by two factors:
one is the thermal force $X_i$ which is the the external cause
describing the environment; the other factor is the transport
coefficients $L_{ij}$ which are the intrinsic causes reflecting
the responsibility of the system driven by the thermal force.
Generally speaking, the entropy variation $\triangle S$ in an unit
time can reflect the evolution of the dissipative non-equilibrium
state. Therefore the Eqs. (\ref{delta s}) and (\ref{flow
expansion}) provide us the basic evolutional information of a
dissipative system theoretically. Inserting Eq. (\ref{flow
expansion}) into Eq. (\ref{delta s}), one could also analyze the
state, or namely the thermal force, of a dissipative system
through the ratio of transport coefficients to the entropy
production which might be inversely obtained through fitting the
experimental data. As for the QGP produced at RHIC, this general
discussion is focalized on the shear viscosity to entropy density
ratio which determines whether it is feasible to describe the hot
and dense medium by the ideal energy-momentum tensor
$T^{\mu\nu}_0$. Besides the shear and bulk viscosity induced by
the velocity gradient, this tensor is also relevant to the thermal
conductivity induced by the energy gradient, which is zero with
Landau and Lifshitz's definition on the hydrodynamic
velocity\cite{Groot}.

In heavy ion physics, the evolution of the QGP density is almost
longitudinal, with velocity gradient along the transverse
direction. Thereby bulk viscosity is usually neglected due to the
domination of  shear viscosity. More proof from recent lattice
simulation favors the ignorance of the bulk viscosity since it is
much smaller than the shear viscous one\cite{Nakamura}. As a
result, the shear viscosity and entropy density ratio becomes a
very important means of understanding the dissipative properties
of the QGP produced in heavy ion collision.

As far as the sQGP is concerned, Ref. \cite{Nakamura} demonstrated
a result from lattice Monte Carlo simulation, pointing out the
ratio of the shear viscosity to the entropy density is smaller
than unit but most probably larger than the universal bound
$1/4\pi$ which is obtained from the gauge theory/gravity
duality\cite{Policastro,Buchel,Kovtun,Maeda} in equilibriate
superstring theory. While one must notice the entropy density in
this ratio is supposed to be an equilibrium one, which is not
exactly the one appears in the Navier-Stocks equation. Some other
estimations\cite{Hirano}, e.g. from the Heisenberg uncertainty
principle, has given a result as $\frac{1}{15}$. Studies on this
strong coupled mechanism are going on since it is desirable to
understand the perfect behavior of the production at RHIC. As to
wQGP(weakly coupled QGP), where the perturbative expansion works,
people knew much more about its viscous properties than those of
the sQGP. The transport coefficients, especially the shear
viscosity, were discussed by many authors in weakly coupled limit
in the kinetics
theory\cite{Hosoya,Gavin,Danielewicz,Oertzen,Baym,Heisenberg,Arnold1,Arnold2,Liu1}
and in the thermal field theory via the Kubo
formulae\cite{jeon1,wang,wang2,Thoma,defu,Aarts,Aarts2,defu2,Jeon,Carrington,Basagoiti,Liu2}.
From the theoretical points of view based on the previous
analysis, the ratio of shear viscosity to entropy density is also
the foundation of understanding the wQGP's dissipative properties.
However most of those investigations mentioned above are usually
focused on the extremely high temperature but zero chemical
potential environment. Further studies on the ratio of shear
viscosity to entropy density at finite density should be carried
on, which is one of our motivations. In addition, as for the
denomination of the ratio, only the equilibrium entropy
density\cite{Policastro,Buchel,Kovtun,Maeda,Blaizot} and the
entropy production in phase transition\cite{Hirano} are discussed.
The entropy production induced by the dissipative forces should be
included. In this paper, considering these two factors we
calculate the shear viscosity of the wQGP at finite chemical
potential and the corresponding entropy production in dissipative
processes self-consistently so as to obtain the viscosity to
entropy density ratio, representing a upper bound in the weak
coupling limit.

The paper is arranged as following: shear viscosity and entropy
density are evaluated in the framework of kinetic theory in Sec. 2
and 3 respectively. Summary and discussions will be presented in
Sec. 4.

\section{Shear viscosity}
In this section, we derive an expression for shear viscosity
$\eta$ of hot QGP at finite chemical potential in leading
logarithm order in the framework of kinetics theory by using
variational approach, following Arnold {\it et al}\cite{Arnold1}
and our previous paper\cite{Liu1}.

\subsection{Formalism and definition}
In a state that is departed not far from the equilibrium, the
energy momentum tensor can be decomposed into ideal and
dissipative parts as
\begin{equation}
T^{\mu\nu}=T^{\mu\nu}_0+\pi^{\mu\nu}=(\epsilon+P)u^\mu u^\nu-Pg^{\mu\nu}+\pi^{\mu\nu}
\end{equation}
where $\epsilon$, $P$, $\pi^{\mu\nu}$ are the energy density,
pressure, and viscous shear stress, respectively. The
four-velocity $u^{\mu}(x)$ in the local rest frame is $(1,0,0,0)$.
In the linear response theory with Landau-Lifshitz convention, the
fluctuation of spacial stress tensor is proportional to the first
order of velocity gradients with neglected bulk viscosity by
defining the coefficient $\eta$ as shear viscosity,
\begin{equation}\label{driving force definition}
\delta\langle\pi_{ij}\rangle=-\eta (\partial_i u_j + \partial_j
u_i-\tfrac{2}{3}\delta_{ij}\partial_k u^k)\equiv-\eta X_{ij},
\end{equation}
where $X_{ij}$ is the so-called driving force.

On one side, in kinetic theory the energy momentum tensor is the
secondary moment of the distribution function,
\begin{equation}
\langle T_{ij}\rangle=\int_\mathbf{p} \frac{p_i p_j}{p^0} \big[g_f
f(t,\mathbf{x};\mathbf{p})+g_{\bar f}{\bar
f}(t,\mathbf{x};\mathbf{p})+g_b b(t,\mathbf{x};\mathbf{p})\big],
\end{equation}
where the momentum space integration $\int_{\mathbf{p}}$ is a
shorthand for $\int\frac{d^3\mathbf{p}}{(2\pi)^3}$, and
$f(x;\mathbf{p})$, $\bar f(x;\mathbf{p})$ and $b(x;\mathbf{p})$
are one particle fermion, anti-fermion and boson distribution
function respectively, with $g_f$, $g_{\bar f}$ and $g_b$ of their
degeneration degrees.

On the other side, when one decomposed the one particle
distribution function as a local equilibrium part plus a fluctuant
part, namely $f_s=n_s+\delta f_s$ where  $n_s$ is the equilibrium
distribution function and $s$ denotes for species of particles,
the fluctuation of the energy moment tensor, which is exactly
contributed by the fluctuation of the stress tensor, is
\begin{equation}\label{viscosity definition}
\delta \langle T_{ij}\rangle=\int_\mathbf{p} \frac{p_i p_j}{p_0}
\big(g_f \delta f+g_{\bar f}{\delta\bar f}+g_b
\delta¡¡b\big)=\delta\langle \pi_{ij}\rangle=-\eta (\partial_i u_j
+ \partial_j u_i-\tfrac{2}{3}\delta_{ij}\partial_k u^k).
\end{equation}

In principle, one could evaluate the shear viscous coefficient as
long as the distribution functions are known. While it is not an
easy task because the one particle distribution function satisfies
the Boltzmann equation of the usual form
\begin{equation}\label{original Boltzmann equation}
\left(\frac{\partial}{\partial t}+\mathbf{\hat
p}\cdot\frac{\partial}{\partial\mathbf{x}}+\mathbf{F_{ext}}\cdot\frac{\partial}{\partial\mathbf{p}}\right)
f_s(t,\mathbf{x};\mathbf{p})=-(\mathcal{C}f_s)(t,\mathbf{x};\mathbf{p}),
\end{equation}
where $\mathbf{F_{ext}}$ is the external field and the
$\mathcal{C}$ is the collision operator. Even in a simple steady
system without external field, this evolution equation contains
both differential and integral terms, where the collision integral
might be rather complicated in different processes. In our paper,
for a upper limit estimation, we just consider $2\rightarrow 2$
elastic collisions to obtain the leading logarithm result,
although for the relaxation and the shear viscosity, the inelastic
scattering might be significant especially in the initial stage of
QGP formation. Under this assumption, the collision operator is
defined as
\begin{eqnarray}
(\mathcal{C}f_s)&&\hspace{-1cm}(\mathbf{x;p})=\frac{1}{2}\int_{\mathbf{p'},\mathbf{k},\mathbf{k'}}\frac{|\mathcal{M}(P,K;P',K')|^2}{(2p_0)(2p'_0)(2k_0)(2k'_0)}
(2\pi)^4\delta^{(4)}(P+K-P'-K')\nonumber\\[0.1cm]
&\times&\left\{f_s(\mathbf{p})f_s(\mathbf{k})[1\pm
f_s(\mathbf{p'})][1\pm f_s(\mathbf{k'})]\right.\nonumber -
\left.f_s(\mathbf{p'})f_s(\mathbf{k'})[1\pm f_s(\mathbf{p})][1\pm
f_s(\mathbf{k})]\right\}\nonumber,
\end{eqnarray},
where $\mathbf{p},\mathbf{k},\mathbf{p'}$ and $\mathbf{k'}$ denote
the momenta of the incoming and outgoing particles respectively.
$|\mathcal{M}|^2$ is the two-body scattering  amplitude. The $1\pm
f_s$ factor is the final state statistical weight, for boson with
the upper sign and for fermion with the down sign.

For convenience, one can rewrite the fluctuation of distribution
function in the form
\begin{equation}\label{delta f}
\delta f_s=n_s(p)[1\pm n_s(p)]\varphi_s(\mathbf{x;p})
\end{equation}

Inserting the decomposed distribution function into the right hand
side of the Boltzmann equation (\ref{original Boltzmann
equation}), one notices that $(\mathcal{C}n_s)(t;\bf{x},\bf{p})=0$
when $n_s$ takes the form of J\"{u}ttner distribution as
\begin{equation}
n_s(t;\bf{x},\bf{p})=\frac{1}{e^{\beta(P\cdot u-\mu_s)}\pm 1}
\end{equation}
with $\mu_s=\pm \mu$ for both fermion and anti-fermion. Notice
that in the local rest frame, this one particle distribution
function is degenerated into the ordinary fermion and boson
distributions,
\begin{eqnarray}
n_f(p)=\frac{1}{e^{\beta (p- \mu)}+1},\ \ \ n_{\bar
f}(p)=\frac{1}{e^{\beta (p+ \mu)}+1},\ \ \ \
n_b(p)=\frac{1}{e^{\beta p}-1}
\end{eqnarray}

Linearizing the collision term, one can obtain
\begin{eqnarray}\label{RHS}
(\mathcal{C}f_a)(\mathbf{x;p})&=&\frac{1}{2}\int_{\mathbf{p'},\mathbf{k},\mathbf{k'}}\sum_{bcd}|{M}_{ab}^{cd}(P,K;P',K')|^2
\times(2\pi)^4\delta^{(4)}(P+K-P'-K')\nonumber\\[0.1cm]
&\times&n_a(p)n_b(k)[1\pm n_c(p')][1\pm
n_d(k')]\nonumber\\[0.1cm]
&\times&\big[\varphi_a(\mathbf{x;p})+\varphi_b(\mathbf{x;k})-\varphi_c(\mathbf{x;p'})-\varphi_d(\mathbf{x;k'})\big
],\nonumber\\
\end{eqnarray}
where $a$ $b$ $c$ $d$ represent the species of the particles and
$|M_{ab}^{cd}(P,K;P',K')|^2$ denotes for
$\frac{|\mathcal{M}_{ab}^{cd}(P,K;P',K')|^2}{(2p_0)(2p'_0)(2k_0)(2k'_0)}$.
The sum in front of the matrix element means all possible
collision processes relevant to the leading-log contribution are
involved and properly treated without double counting or
multi-counting.

As for the left hand side of the Boltzmann equation, the gradients
acting on $\delta f_s$ give higher order in the  departure from
equilibrium, so that in the first order approximation, only $n_s$
should be considered on this side, namely,
\begin{equation}\label{LHS}
LHS=\beta n_s(p)[1\pm
n_s(p)]I_{ij}(\hat{\mathbf{p}})X^s_{ij}(\mathbf{x})\\
\end{equation}
with
$I_{ij}(\mathbf{\hat{p}})=\frac{1}{2}(\hat{p_i}\hat{p_j}-\frac{1}{3}\delta_{ij})$,
where the time derivative and the the external field terms vanish
for viscosity and the spatial tensor $X^s_{ij}(\mathbf{x})$ is
just the driving force defined in Eq. (\ref{driving force
definition}).

Comparing  both sides of the Boltzmann equation, one finds that
$\varphi(\mathbf{x;p})$ on the right hand side must have the same
angular dependence as the driving term, i.e.
\begin{equation}\label{phi function}
\varphi_s(\mathbf{x;p})=\beta^2 I_{ij}(\mathbf{\hat{p}})
X_{ij}(\mathbf{x})\chi_s(p)
\end{equation}
where $\chi_s(p)$ is rotationally invariant function depending
only on the amplitude of momentum.

With equations (\ref{RHS}),(\ref{LHS}) and (\ref{phi function}),
one cancels the driving field on both sides and recasts the
Boltzmann equation into
\begin{equation}\label{recasted B equation}
S^s_{ij}(\mathbf{p})=(\mathcal{C}\chi^s_{ij})(\mathbf{p})
\end{equation}
where
\begin{eqnarray}
S^s_{ij}(\mathbf{p})&\equiv& -Tp\ n_s(p)[1\pm
n_s(p)]I_{ij}(\mathbf{\hat{p}}),\\
\chi^s_{ij}(\mathbf{p})&\equiv&I_{ij}(\hat{\mathbf{p}})\chi_s(p).
\end{eqnarray}

To solve this equation, we introduced the variational approach
following the basic steps in reference\cite{Arnold1}. First, one
defines an inner product as
\begin{equation}
(f,g)\equiv \beta^3\int_\mathbf{p}f(\mathbf{p})g(\mathbf{p}).
\end{equation}
Then taking product on both sides of the Boltzmann Eq.
(\ref{recasted B equation}) and defining the functional
\begin{equation}
Q[\chi]\equiv(\chi_{ij},S_{ij})-\tfrac{1}{2}(\chi_{ij},\mathcal{C}_{ij})
\end{equation}
with explicit form of each term as
\begin{equation}
(\chi_{ij},S_{ij})=-\beta^2\sum\limits_a \int_{\mathbf{p}}\
\mathbf{p}\ f^a_0(\mathbf{p})[1\pm f^a_0(\mathbf{p})] \chi^a (p),
\end{equation}
\begin{eqnarray}\label{collision term}
(\chi_{ij},\mathcal{C}\chi_{ij})&=&\frac{\beta^2}{8}\int_{\mathbf{p},\mathbf{k},\mathbf{k},\mathbf{k'}}
\sum\limits_{abcd}|{M}^{ab}_{cd}|^2\
(2\pi)^4\delta^{(4)}(P+K-P'-K')\nonumber\\[0.3cm]
&\times& n_a(\mathbf{p})n_b(\mathbf{k})[1\pm
n_c(\mathbf{p'})][1\pm n_d(\mathbf{k'})]
\nonumber\\[0.2cm]
&\times&\left
[\chi_{ij}^a(\mathbf{p})+\chi_{ij}^b(\mathbf{k})-\chi_{ij}^c(\mathbf{p'})-\chi_{ij}^d(\mathbf{k'})\right
]^2,
\end{eqnarray}
one may relate the maximal value of Q with the viscous coefficient
through Eqs. (\ref{viscosity definition}), (\ref{delta f}) and
(\ref{phi function}) with the fact that the maximum value of
$Q[\chi]$ occurs when $\chi(p)$ satisfies Eq. (\ref{recasted B
equation}), i.e.,
\begin{equation}
\eta=\tfrac{2}{15}Q_{max}\big|_{\chi=\chi_{max}}.
\end{equation}

\subsection{Collision terms}
\begin{figure}
 \begin{center}
   \resizebox{11cm}{!}{\includegraphics{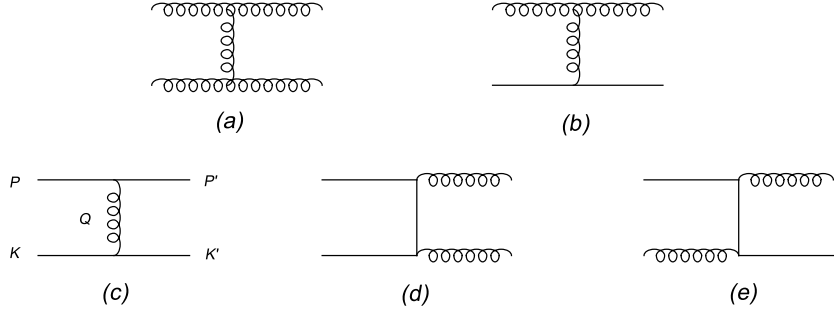}}
 \end{center}
   \caption{\label{fig1}The possible processes which contribute to the leading-log in the collision term in QCD plasma.
   The solid line is for quark and the wiggly line is for gluon. The matrix elements arise from all the five diagrams
   are $16g^4 d_A
C_A^2\left(3-\frac{su}{t^2}-\frac{st}{u^2}-\frac{tu}{s^2}\right)$,
$8g^4d_F C_F C_A\left(\frac{s^2+u^2}{t^2}\right)$,
$8g^4\frac{d_F^2
C_F^2}{d_A}\left(\frac{s^2+u^2}{t^2}+\frac{s^2+t^2}{u^2}\right)$,
$8g^4 d_F C_F^2\left(\frac{t}{u}+\frac{u}{t}\right)$ and $-8g^4
d_F C_F^2\left(\frac{s}{u}+\frac{u}{s}\right)$, respectively, with
$d_A=8$, $d_F=C_A=3$ and $C_F=\tfrac{4}{3}$ for SU(3) group.}
\end{figure}

The collision terms of the Boltzmann equation are associated with
different interaction processes. In the leading logarithm order,
only five diagrams contribute, which are presented in
Fig.(\ref{fig1}).

Before starting calculation of these collision terms , some
arguments and kinematic relations should be manifested first.
\begin{itemize}
\item Approximations

Two approximations have been employed in our further evaluation.
First, the high temperature approximation. We assume the
temperature $T$ of the QGP is extremely high which is the only
'hard scale' in the system, with other quantities like chemical
potential $\mu$, thermal mass etc. indicated by $gT$, are  much
less than the hard scale in weakly coupled theory. With this high
temperature approximation, the fermion(anti-fermion) distribution
function can be expanded in terms of $\mu/T$, neglecting the
thermal mass term directly\cite{Liu1}. The second important
approximation is the forward scattering approximation, i.e. the
momentum transfer between the incident particles $q\sim gT$ is
rather small so that the momentum difference between the incoming
and outgoing particles on the same interaction vertex can be
ignored in the distribution functions.

\item Kinematics

We mark the momenta of the particles as shown in
Fig.\ref{fig1}(c). The three usual Mandelstam variables are
defined as $s=(P+K)^2$, $t=(P-P')^2$ and $u=(P-K')^2$. With the
forward scattering approximation, the incident angle between
$\mathbf{p}$ and $\mathbf{k}$ can be described by $\theta$ and
$\phi$, which are the angle between $\mathbf{p}$ and $\mathbf{q}$
and the angle between the $\mathbf{p}\mathbf{q}$ plane and
$\mathbf{p'}\mathbf{q}$ plane,
\begin{equation}\label{kinematics}
\cos\theta_{pk}=1+(1-\cos^2\theta)(1-\cos\phi).
\end{equation}
\end{itemize}

Now let us turn back to the Eq. (\ref{collision term}). Performing
the integral over $d\mathbf{k'}$ with the help of
$\delta^3(\mathbf{p}+\mathbf{k}+\mathbf{p'}+\mathbf{k'})$, one may
transform the equation into
\begin{eqnarray}\label{collision term 2}
(\chi_{ij},\mathcal{C}\chi_{ij})&=&\frac{\beta^3}{(4\pi)^6}\int_0^\infty
dq \int_{-q}^q d\omega \int_0^\infty dp \int_0^\infty dk
\int_0^{2\pi}d\phi\nonumber\\[0.1cm]
&&\hspace{-0.4cm}\sum\limits_{abcd}|\mathcal{M}^{ab}_{cd}|^2
n_a(p)n_b(k)[1\pm
n_c(p)][1\pm n_d(k)]\nonumber \\
&\times&\left
[\chi_{ij}^a(\mathbf{p})+\chi_{ij}^b(\mathbf{k})-\chi_{ij}^c(\mathbf{p'})-\chi_{ij}^d(\mathbf{k'})\right
]^2.
\end{eqnarray}
where $\omega$ is a dummy integration variable first introduced by
Baym\cite{Baym} with $p'=p+\omega$ and $k'=k-\omega$.

It is easy to discover that the integrand of Eq. (\ref{collision
term 2}) is composed by three parts: the matrix element, the
distribution functions and the $\chi$ term. According to the
exchanged particle, the $\chi$ term can be sorted into two
classes: Fig.\ref{fig1}(a-c) belong to one category where the
on-shell particles interact with each other through exchanging a
boson, namely, they have the same species of incoming and outgoing
particles on one interaction vertex; Fig.\ref{fig1}(d) and (e)
belong to another category where they contain off-shell fermions,
i.e., different species of on-shell particles are bounded to the
same interaction vertex. For the first category, the $\chi$ term
contributes a small $q^2$ in the forwarding scattering
approximation, which softens the infrared singularityand gives the
leading-log form of viscosity\cite{Arnold1}. For the two on-shell
line of quarks, the $\chi$ term thus is specified as
\begin{equation}
\left[\chi^q_{ij}(\mathbf{p'})-\chi^q_{ij}(\mathbf{p})\right]^2\approx\left[\mathbf{q}\cdot\nabla\chi^q_{ij}(\mathbf{p})\right]^2
=\omega^2[\chi^q(p)']^2 +3\frac{q^2-\omega^2}{p^2}[\chi^q(p)]^2.
\end{equation}
where $\chi^q(p)'=d\chi^q(p)/dp$. We assumed that the quark and
anti-quark have the same departure behavior from the equilibrium
state which is denoted by $\chi^q$, since nothing but the viscous
process is involved. The same trick will be performed on $\chi^g$,
the gluon $\chi$ function, and leads to similar result. One can
prove that the interference term like
$[\chi^q(\mathbf{p})-\chi^q(\mathbf{p'})][\chi^g(\mathbf{k})-\chi^g(\mathbf{k'})]$
vanishes when carrying out the $d\omega$ and $d\phi$ integration.
Following the above discussion, we can list the all the $\chi$
terms for different diagrams in Table\ref{chi term}, where the
only survived formats are $(\chi^q-\chi^q)^2$, $(\chi^q-\chi^g)^2$
and $(\chi^g-\chi^g)^2$.
\begin{table}
\begin{eqnarray}\renewcommand{\arraystretch}{2}
\begin{array}{|c||c|}\hline
\mbox{Processes}                             &    \mbox{$\chi$ functions} \\
\hline\hline \mbox{Fig.\ref{fig1}(a)}        &
[\chi^g(\mathbf{p})-\chi^g(\mathbf{p'})]^2+[\chi^g(\mathbf{k})-\chi^g(\mathbf{k'})]^2\\
\hline \mbox{Fig.\ref{fig1}(b1)}             &  [\chi^g(\mathbf{p})-\chi^g(\mathbf{p'})]^2+[\chi^q(\mathbf{k})-\chi^q(\mathbf{k'})]^2\\
\mbox{Fig.\ref{fig1}(b2)}                    &  [\chi^q(\mathbf{p})-\chi^q(\mathbf{p'})]^2+[\chi^g(\mathbf{k})-\chi^g(\mathbf{k'})]^2\\
\hline\mbox{Fig.\ref{fig1}(c)}               &  [\chi^q(\mathbf{p})-\chi^q(\mathbf{p'})]^2+[\chi^q(\mathbf{k})-\chi^q(\mathbf{k'})]^2\\
\hline \mbox{Fig.\ref{fig1}(d)(e)}            &
  [\chi^q(\mathbf{p})-\chi^g(\mathbf{p'})]^2+[\chi^q(\mathbf{k})-\chi^g(\mathbf{k'})]^2\\
\hline
\end{array}
\nonumber
\end{eqnarray}
\caption{\label{chi term}The $\chi$ terms for five diagrams. The
Fig.\ref{fig1}(b) has two sets of $\chi$ functions be
cause it involves different channels which bring on different
momentum dependence of $\chi^q$ and $\chi^g$.}
\end{table}

There are two facts that should be clarified. One is that all
possible channels for one diagram are not to be added directly. In
fact for each channel, one must perform a convolution integral for
the exact $\chi$ term from the very channel with the corresponding
distribution function. The other fact to be noticed is the
relationship between the Mandelstam variables under the forward
scattering approximation. For instance, as to the massless
on-shell particles, $s=(P+K)^2=2P\cdot K\approx -(P-K')^2=-u$ in
u-channel.Similarly, one has $s\approx -t$ in t-channel. With this
convenience, combining with Eq. (\ref{kinematics}) one finds out
that all the matrix elements reduce to only two forms,
\begin{eqnarray}
&&\left(\frac{s^2+t^2}{u^2}\right)_{u-channel}=\left(\frac{s^2+u^2}{t^2}\right)_{t-channel}\approx\frac{8p^2k^2}{q^4}(1-\cos\phi)^2,
\\[0.3cm]
&&\left(\frac{t}{u}\right)_{u-channel}=\left(\frac{u}{t}\right)_{t-channel}\approx\frac{2pk}{q^2}(1-\cos\phi).
\end{eqnarray}

The last part of collision integrand is the distribution function,
which is the main distinguished feature for each diagram and
channel. Analyzing each possible process carefully, one can obtain
the distribution function term for each diagram in Fig. \ref{fig1}
which is listed in Tabel \ref{distribution function}.
\begin{table}
\begin{eqnarray}\renewcommand{\arraystretch}{2}
\begin{array}{|c||l|}
\hline \mbox{Processes}    &  \mbox{\hspace{3cm}Distribution functions}\\
\hline\hline
\mbox{Fig.\ref{fig1}(a)}   & 2n_b(p) n_b(k)[1+n_b(p)][1+n_b(k)]\\
\hline
\mbox{Fig.\ref{fig1}(b1)}  & 2N_f \big\{n_f(k)[1-n_f(k)]+n_{\bar f}(k)[1-n_{\bar f}(k)]\big\} n_b(p)[1+n_b(p)]\\
\mbox{Fig.\ref{fig1}(b2)}  & 2N_f n_b(k)[1+n_b(k)]\big\{n_f(p)[1-n_f(p)]+n_{\bar f}(p)[1-n_{\bar f}(p)]\big\}\\
\hline
\mbox{Fig.\ref{fig1}(c)}   & 2N_f^2 \big\{n_f(p) n_f(k)[1-n_f(p)][1-n_f(k)]+n_{\bar f}(p) n_f(k)[1-\bar n_{\bar f(k)}][1-n_f(k)]\\
                           & +n_f(p) n_{\bar f}(k)[1-n_f(p)][1-n_{\bar f}(k)]+n_{\bar f}(p) n_{\bar f}(k)[1-n_{\bar f}(p)][1-n_{\bar f}(k)]\big\}\\ \hline
\mbox{Fig.\ref{fig1}(d+e)} & 2N_f \big\{n_f(p) n_b(k)[1-n_f(k)][1+n_b(p)]+n_{\bar f}(p) n_b(k)[1- n_{\bar f(k)}][1+n_b(p)]\\
                           & +n_b(p) n_f(k)[1-n_f(p)][1+ n_b(k)]+n_b(p) n_{\bar f}(k)[1-n_{\bar f(k)}][1+n_b(k)]\big\}\\
\hline
\end{array}\nonumber
\end{eqnarray}
\caption{\label{distribution function}The distribution function
terms for the five processes, where $N_f$ is the quark flavor. The
factors in front of the distribution functions are the freedom of
degeneration, which are related to the distinguished reaction
channels. For example, $q\bar q\leftrightarrow q\bar q$ appears
$4N_f$ times in the sum over species. }
\end{table}

Inserting the expressions in the three tables into Eq.
(\ref{collision term 2}) and carrying on the integration over
$dk$, $d\phi$ and $d\omega$, one can finally obtain the full
collision term,
\begin{eqnarray}\label{right side}
(\chi_{ij},\mathcal{C}\chi_{ij})=&&\hspace{-0.8cm}(\chi_{ij},\mathcal{C}\chi_{ij})^{(a)}+(\chi_{ij},\mathcal{C}\chi_{ij})^{(b1)}
+(\chi_{ij},\mathcal{C}\chi_{ij})^{(b2)}\nonumber \\
&+&(\chi_{ij},\mathcal{C}\chi_{ij})^{(c)}+(\chi_{ij},\mathcal{C}\chi_{ij})^{(d+e)}
\end{eqnarray}
where
\begin{eqnarray}
(\chi_{ij},\mathcal{C}\chi_{ij})^{(a)}=\frac{\alpha_s^2 d_A
C_A^2}{3\pi}\int^T_{\alpha_s T}\frac{dq}{q}\int^\infty_0 dp\
n_b(p)\big[1+n_b(p)\big]\big\{p^2\left[\chi^g(p)'\right]^2+6\left[\chi^g(p)\right]^2\big\},\nonumber\\
\end{eqnarray}
\begin{eqnarray}
(\chi_{ij},\mathcal{C}\chi_{ij})^{(b1)}=&&\hspace{-0.9cm}\frac{2\alpha_s^2
d_F N_f C_F C_A}{3\pi}\int^T_{\alpha_s T}\frac{dq}{q}\int^\infty_0
dp\
\big\{p^2\left[\chi^q(p)'\right]^2+6\left[\chi^q(p)\right]^2\big\}\nonumber\\
&\times&\big\{n_f(p)\big[1-n_f(p)\big]+n_{\bar f}(p)\big[1-n_{\bar
f}(p)\big]\big\},
\end{eqnarray}
\begin{eqnarray}
(\chi_{ij},\mathcal{C}\chi_{ij})^{(b2)}=&&\hspace{-0.9cm}\frac{2\alpha_s^2
d_F N_f C_F
C_A}{3\pi}\left(1+\frac{3}{\pi^2}\frac{\mu^2}{T^2}\right)\int^T_{\alpha_s
T}\frac{dq}{q}\int^\infty_0
dp\ \big\{n_b(p)\big[1+n_b(p)\big]\big\}\nonumber\\
&\times&\big\{p^2\left[\chi^g(p)'\right]^2+6\left[\chi^g(p)\right]^2\big\},
\end{eqnarray}
\begin{eqnarray}
(\chi_{ij},\mathcal{C}\chi_{ij})^{(c)}=&&\hspace{-0.9cm}\frac{2\alpha_s^2
(d_F N_f C_F)^2}{3
d_A\pi}\left(1+\frac{3}{\pi^2}\frac{\mu^2}{T^2}\right)\int^T_{\alpha_s
T}\frac{dq}{q}\int^\infty_0
dp\ \big\{p^2\left[\chi^q(p)'\right]^2+6\left[\chi^q(p)\right]^2\big\}\nonumber\\
&\times&\big\{n_f(p)\big[1-n_f(p)\big]+n_{\bar f}(p)\big[1-n_{\bar
f}(p)\big]\big\},
\end{eqnarray}
\begin{eqnarray}
(\chi_{ij},\mathcal{C}\chi_{ij})^{(d+e)}&&\hspace{-0.9cm}=\frac{2\alpha_s^2
d_F N_f C_F^2 \beta}{8 d_A\pi^3}\int^T_{\alpha_s
T}\frac{dq}{q}\int^\infty_0
dp\ p \left[\chi^q(p)-\chi^g(p)\right]^2\nonumber\\
 &\times&\left\{n_f(p)\big[1+n_b(p)\big]\left(\frac{\pi^2}{8}-0.616\frac{\mu}{T}+\frac{\mu^2}{8T^2}\right)\right. \nonumber\\
&&+n_{\bar f}(p)\big[1+n_b(p)\big]\left(\frac{\pi^2}{8}+0.616\frac{\mu}{T}+\frac{\mu^2}{8T^2}\right) \nonumber\\
&&+b_f(p)\big[1-n_f(p)\big]\left(\frac{\pi^2}{8}+0.616\frac{\mu}{T}+\frac{\mu^2}{8T^2}\right) \nonumber\\
&&+\left.n_b(p)\big[1-n_{\bar
f}(p)\big]\left(\frac{\pi^2}{8}-0.616\frac{\mu}{T}+\frac{\mu^2}{8T^2}\right)\right\},
\end{eqnarray}
with $\alpha_s$ as the fine structure constant. We replace the
limits of $dq$ integration by the hard and the soft scale $T$ and
$\alpha_s T$ due to the leading-log treatment, which includes the
self-energy effect of the exchange line\cite{Arnold1}.

\subsection{Variational method}

So far we have obtained the right hand side of the Boltzmann
equation, and the other side can be written down directly as
\begin{eqnarray}\label{left side}
(\chi_{ij},S_{ij})=-\frac{\beta^2}{\pi^2}\int_0^\infty dp
\hspace{-0.6cm}&& p^3\Big\{ d_F N_f \big\{
n_f(p)[1-n_f(p)]+n_{\bar f}(p)[1-n_{\bar f}(p)]\big\}\chi^q(p)\nonumber\\
&&+d_A n_b(p)[1+n_b(p)]\chi^g(p)\Big\}
\end{eqnarray}

In this subsection, we expand the trial function $\chi(p)$ by a
finite set of basis with independent variational parameters to
maximize the functional $Q[\chi]$.

As we have known that in the case of viscosity, the $\chi$
function has two components, $\chi^q$ and $\chi^g$,
\begin{equation}
\chi(p)= \left(
\begin{array}{c}
\chi^g(p) \\
\chi^q(p)
\end{array}
\right).
\end{equation}
Expanding the two components by the same basis
\begin{equation}\label{basis expansion}
\chi^g(p)=\sum\limits_{m=1}^N a_m \phi_m(p),\ \ \
\chi^q(p)=\sum\limits_{m=1}^N a_{N+m} \phi_m(p),
\end{equation}
one could read out the basis-set components of $\tilde S$ and
$\tilde{{C}}$
\begin{equation}
(S_{ij},\chi_{ij})=\sum\limits_m a_m\tilde S_m, \ \ \
(\chi_{ij},\mathcal{C}\chi_{ij})=\sum\limits_{mn}a_m\
\tilde{C}_{mn}\ a_n.
\end{equation}
When the Boltzmann equation is satisfied, the coefficients in
front of the bases will be expressed as $a=\tilde{{C}}^{-1}\tilde
S$. Accordingly, the shear viscosity becomes
\begin{equation}
\eta=\tfrac{2}{15}{Q}_{max}=\tfrac{1}{15}a\cdot\tilde{S}=\tfrac{1}{15}\tilde{S}^t\tilde{C}^{-1}\tilde{S}
\end{equation}

With the natural one function ansatz $\phi_{1}(p)=p^2/T$,  one can
evaluate the integral of Eqs. (\ref{right side}) and (\ref{left
side}) analytically,
\begin{eqnarray}
\tilde S&=&-\frac{\beta^3}{\pi^2}\!\!\int^\infty_0 \!\! dp\ p^5
\left(
   \begin{array}{c}
   d_A n_b(p)[1+n_b(p)]\\[0.2cm]
   d_F N_f\left\{ n_f(p)[1-n_f(p)]+n_{\bar f}(p)[1-\bar
    n_f(p)]\right\}
\end{array}
\right)\nonumber\\[0.3cm]
&=&-\frac{120\zeta(5)T^3}{\pi^2} \left(
\begin{array}{c}
d_A\\[0.3cm]
\frac{15}{8}d_F
N_f\left(1+\frac{2\zeta(3)}{5\zeta(5)}\frac{\mu^2}{T^2}\right)
\end{array}
\right)
\\[1cm]
\mathcal{\tilde{C}}&=&\frac{16\pi^3\alpha_s^2 T^3 \ln \alpha_s
^{-1}}{9d_A}\left[ \left(
\begin{array}{cc}
\hspace{-2cm}d_A C_A\big[d_A C_A &0 \\[0.2cm]
\hspace{1cm}+N_f d_F C_F(1+\tfrac{3}{\pi^2}\frac{\mu^2}{T^2})\big]
&\\[0.3cm]
\ &\frac{7}{4}N_f d_F C_F\big[d_A
C_A(1+\frac{30}{7\pi^2}\frac{\mu^2}{T^2})\\[0.3cm]
0 &\hspace{1cm}+N_f d_F
C_F(1+\frac{51}{7\pi^2}\frac{\mu^2}{T^2})\big]
\end{array}\right)\right.\nonumber\\[0.5cm]
&&\hspace{1cm}+\left.\frac{9\pi^2}{128}N_f d_F C_F^2
d_A\left(1-0.013\frac{\mu^2}{T^2}\right)\left(
\begin{array}{cc}
1&-1\\[0.3cm]
-1&1
\end{array}
\right ) \right].
\end{eqnarray}
 where $\zeta(x)$ is the Riemann $\zeta$-function. Then one can
compute the coefficient vector and the shear viscosity of 2-flavor
QGP at finite temperature and chemical potential as,
\begin{equation}\label{a solution}
a=\left(
\begin{array}{c}
a_1\\
a_2
\end{array}
\right) = (g^4\ln \alpha_s ^{-1})^{-1}\left(
\begin{array}{c}
-3.281+0.940\frac{\mu^2}{T^2}  \\
-6.792+0.199\frac{\mu^2}{T^2}
\end{array}
\right)
\end{equation}
and
\begin{equation}\label{viscosity of QGP}
\eta\approx\frac{1.09T^3}{\alpha_s^2\ln
\alpha_s^{-1}}\left(1+0.25\frac{\mu^2}{T^2}\right)
\end{equation}
which recovers the result in Ref. \cite{Arnold1} when the chemical
potential vanishes.

\section{Entropy production in viscous process}

From the non-equilibrium thermodynamical points of view, the
entropy $S$ symbols a scale that how far the system is apart from
the equilibrium state where the entropy takes the maximum value.
Other than the viscosity which reflects the intrinsic property of
a system, the entropy depends further on the environment the
system resides in, or the velocity gradient in our case. It is the
velocity gradient that determined the evolution from
non-equilibrium to equilibrium in a given system. Unfortunately,
we know little about the velocity distribution which depends on
the experimental environment and should be obtained by fitting
data. In Sec. 4, a maximum estimation of the velocity gradient
will be employed so as to demonstrate a maximum entropy production
which leads to an upper bound of viscosity to entropy density
ratio.

We introduce two different schemes to calculate the entropy
production and find out that their results are comparable
qualitatively.

\subsection{Scheme I}
In relativistic kinetic theory, the entropy density of a system is
defined as
\begin{equation}\label{entropy kinetics}
s=-\sum\limits_s\int_\mathbf{p}g_s\Big\{f_s(\mathbf{x;p})\ln
f_s(\mathbf{x;p})\mp \big[1\pm f_s(\mathbf{x;p})\big]\ln \big[1\pm
f_s(\mathbf{x;p})\big]\Big\}
\end{equation}
where $g_s$ is the degeneration degree of each particle species.

Decomposing the distribution functions and inserting Eq.
(\ref{delta f}) into Eq. (\ref{entropy kinetics}), one finds that
only the second order in $\varphi$ survives while the first order
in $\varphi_s$ vanishes due to the rotational invariance of
$n_s(p)$. The entropy density thus becomes
\begin{equation}\label{entropy}
s=s_0-\int_\mathbf{p}\sum\limits_s g_s n_s(p)\big[1\pm
n_s(p)\big]\varphi_s^2(\mathbf{x;p})
\end{equation}
where $s_0$ is the entropy density in the equilibrium state,
\begin{eqnarray}\label{equilibrium entropy}
s_0&=&-\int_\mathbf{p}\Big\{g_f\big[n_f \ln n_f + (1-n_f)\ln
(1-n_f)\big]+g_{\bar f}\big[n_{\bar f} \ln n_{\bar f}
\nonumber\\
&&+(1-n_{\bar f})\ln (1-n_{\bar f})\big]+ g_b\big[n_b
\ln n_b-(1+n_b)\ln(1+n_b)\big] \Big\}\nonumber\\
&=&16.22T^3\left(1+0.123\frac{\mu^2}{T^2}\right)
\end{eqnarray}
with $g_f=g_{\bar f}=2N_f d_F$ and $g_b=2d_A$ and the integral
evaluated by expanding the fermion distribution functions in terms
of small $\mu/T$. It is clear that the second term of Eq.
(\ref{entropy}) denotes for the departure of the entropy density
from its maximum value, i.e., the entropy production. As general,
the non-equilibrium entropy density has the structure of
$s=s_0-\triangle s$, as we expect.

In this scheme we are going to directly evaluate the second term
of Eq. (\ref{entropy}) by adopting the solutions of the Boltzmann
equation in Eq. (\ref{a solution}), leaving only the velocity
gradient to be estimated. Inserting expression(\ref{phi function})
into Eq. (\ref{entropy}) with the solved $\chi(p)$ in the
variational approach, we can obtain
\begin{eqnarray}\label{entropy production 1}
s&=&s_0-\frac{\beta^6}{120\pi^2}\int_0^\infty dp\
p^6\left\{g_f\big[n_f(1-n_f)+n_{\bar f}(1-n_{\bar f})\big ]a_1^2\right.\nonumber\\
&&\hspace{4cm}+\left.g_b
n_b(1+n_b)a_2^2\right\}\left(\frac{\partial u_z}{\partial
x}\right)^2 \\
&=& s_0-\frac{9.75T}{(\alpha_s^2\ln
\alpha_s^{-1})^2}\left(1+0.284\frac{\mu^2}{T^2}\right)\left(\frac{\partial
u_z}{\partial x}\right)^2.
\end{eqnarray}

It's obvious that the second term of Eq. (\ref{entropy production
1}) is the entropy production relative to viscous process when the
system evolving to its equilibrium state, which depends on the
kinetic parameter $\alpha_s$, the thermodynamic parameters $T$
$\mu$ and the transport environment $\partial u_z/\partial x$. To
understand further on their physical meaning, we introduce a
second scheme to evaluate the entropy produced in the viscous
process.

\subsection{Scheme II}

From Scheme I, one can see the non-equilibrium entropy density is
obtained by subtracting the entropy production from the
equilibrium entropy density, which is denoted by
\begin{equation}
s=s_0-\int dX \partial_\mu s^\mu=s_0-\int dt \partial_t s^0
\end{equation}
where $s^\mu=s u^\mu$ is the four entropy flow. The last equality
is based on the description of longitudinal expansion model in
local rest frame.

As we stated in Sec. 1, the entropy production could be obtained
by the product of the driving force and corresponding
thermodynamic flux. Generally speaking, the entropy produced in
unit phase space $\partial_\mu s^\mu$ takes a bilinear form: it is
the sum of all kinds of transport flows, each multiplied by a
characteristic thermodynamic driving force. Furthermore, under the
linear law, the transport flow is proportional to the the driving
force with the coefficient of viscosity, conductivity, etc.
Therefore, the universal expression for the entropy production
should be the combination of squared driving forces coming from
various transport processes, weighted by the corresponding
transport coefficients. In our case, we would like to know the
entropy production just from the viscous process, which
is\cite{Hosoya}
\begin{equation}\label{ep in scheme2}
\partial_\mu  s^\mu=\partial_t  s^0=\frac{\eta}{T}
X_{ij}^2=\frac{2\eta}{T}\left(\frac{\partial u_z}{\partial
x}\right)^2
\end{equation}
where $X_{ij}$ is defined in expression(\ref{driving force
definition}). Noticing that the integration variable $t=\tau \cosh
y$, where $\tau$  describes the time scale of system departure,
one could replace it by the relaxation time $\tau_{\eta}$ for the
maximum estimation. With this approximation, one can perform the
integration on the pseudorapidity plateau,
\begin{eqnarray}\label{e1}
s&\leqslant& s_0-\frac{4\eta\tau_{\eta}}{T R_A^2}\int_1^{\cosh
y_0}
(\sinh y)^2 d(\cosh y)\\
&=& s_0-\frac{1.875T}{(\alpha_s^2\ln
\alpha_s^{-1})^2}\left(1+0.25\frac{\mu^2}{T^2}\right)\left(\frac{\partial
u_z}{\partial x}\right)^2.
\end{eqnarray}
where $y_0=2$ is the edge value of the pseudorapidity plateau and
$\tau_{\eta}$ will be calculated in the appendix.

We argue that this estimation is consistent, for the relaxation
time defined in the Boltzmann equation describes the one particle
property in the proper reference frame. From Eq. (\ref{ep in
scheme2}) one can see clearly that the entropy produced in viscous
process depends not only on the intrinsic property $\eta$ but also
on the exterior transport environment of the system. These
internal and external factors determine the dependence of the
entropy density on the kinetic and thermodynamic parameters, and
also the velocity gradient.

\section{The ratio of $\eta$ to $s$}

Now we are going to turn to a significant quantity $\eta/s$ which
has been mentioned in the first section. In a pure viscous
transport process, the state deviating from the equilibrium is
solely stimulated by the spacial derivative of velocity field
$\partial_i u_j$. For a rough estimation of the velocity gradient
for a longitudinal Lorentz boost invariant system, we adopt
Brioken's assumption $u^\mu=x^\mu/\tau$\cite{Brjoken}, where
$x^\mu$ is the space-time point and $\tau$ is the proper time. In
this scenario, the shear velocity is just along longitudinal
direction $\hat{\mathbf{z}}$ which is perpendicular to the
velocity gradient direction $\hat{\mathbf{x}}$. Thereby
$\partial_i u_j$ has just one non-zero component $\partial u_z
/\partial x$. Noticing that in the light-cone coordinate,
$u_z=x_z/\tau=\sinh y$ where $y$ denotes the pseudo-rapidity and
the maximum of velocity gradient is $\partial u_z/\partial
x\leqslant u_z/R_A=\sinh y/R_A$ when $y$ takes the edge value of
the pseudo-rapidity plateau. Here $R_A$ is the nucleus radius and
 the system is considered rotational invariant around the z-axis.
In Au+Au $200GeV$ collision at RHIC, the pseudo-rapidity plateau
is from $-2\sim 2$\cite{star } in the most central collision, and
the radius of gold nucleus is about $35.381GeV^{-1}$.

With this maximal velocity gradient estimation, the ratio of
viscosity to entropy density obtained in the two schemes are
\begin{eqnarray}\label{ratio1}
\left(\frac{\eta}{s}\right)_{\mbox{\tiny
I}}\geqslant\frac{\eta}{s_0}\left[1-\frac{0.013GeV^2}{(T
\alpha_s^2\ln \alpha_s^{-1})^2
}\left(1+0.16\frac{\mu^2}{T^2}\right)\right]^{-1},
\end{eqnarray}
\begin{eqnarray}\label{ratio2}
\left(\frac{\eta}{s}\right)_{\mbox{\tiny
II}}\geqslant\frac{\eta}{s_0}\left[1-\frac{0.23GeV^2}{(T
\alpha_s^2\ln \alpha_s^{-1})^2
}\left(1+0.12\frac{\mu^2}{T^2}\right)\right]^{-1},
\end{eqnarray}
respectively. These two results are qualitatively comparable for
they have exactly the same dependence on dynamic and thermodynamic
parameters except some differences on the factors.

We manage to demonstrate the basic features of the dependence of
the ratio on the thermal parameters by adopting the result from
scheme I. As their chemical potential dependence is quite clear,
we here just present their variation on the temperature. To
isolate the temperature dependence, we use a simple form of the
running coupling constant\cite{Nakamura}
\begin{equation}\label{coupling constant}
\alpha_s=\frac{2\pi}{11}\left(\ln\frac{4T}{1.5T_c}\right)^{-1},
\end{equation}
where $T_c$ is the critical temperature of QGP phase transition.
We assume the chemical potential has negligible effect on the
constant since it is small compared to the temperature above
$T_c$. Inserting Eq.(\ref{coupling constant}) into
Eq.(\ref{ratio1}), we can obtain Fig. 2 with $T_c=175MeV$ and
$\mu=46MeV$.

\begin{figure}
 \begin{center}
   \resizebox{8cm}{!}{\includegraphics{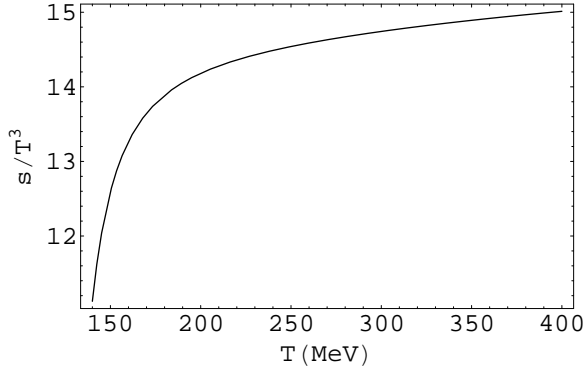}}
 \end{center}
   \caption{The shear viscosity to entropy density ratio of 2-flavor QGP.}
\end{figure}
It should be noticed here that in Fig. 2 the temperature
dependence of the ratio has been extrapolated into the strongly
coupled region where $\alpha_s>1$ to see where is the physical
boundary and the whole variety of the ratio in the physical
region. One may find that the structure of the viscosity to
entropy ratio is rather complicated due to the appearance of two
singularities, which are contributed by the logarithm and the
denominator of the enhancement factor. To distinguish the physical
curve from the complex structure, we naturally introduce a
physical criterion: both the ratio in equilibrium and the
enhancement factor, which are presented in Figs. 2 and 3, should
be positive.
\begin{figure}
 \begin{center}
   \resizebox{8cm}{!}{\includegraphics{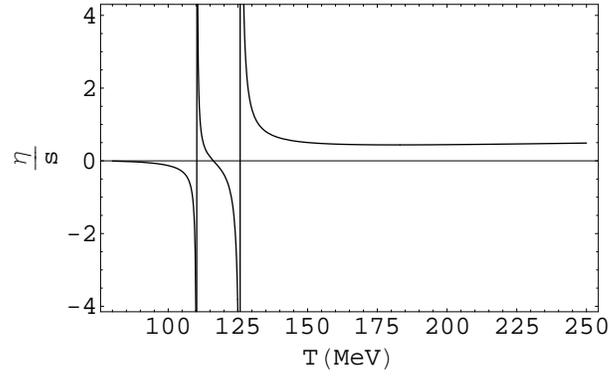}}
 \end{center}
   \caption{Ratio of shear viscosity to entropy density in equilibrium.}
\end{figure}

\begin{figure}\label{fig enhancement factor}
 \begin{center}
   \resizebox{8cm}{!}{\includegraphics{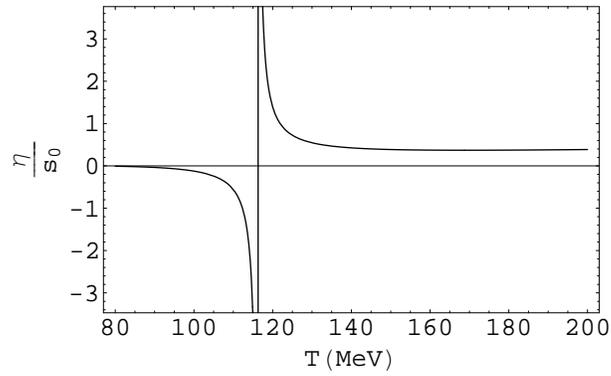}}
 \end{center}
   \caption{Enhancement factor.}
\end{figure}

It is clear in Figs. 2 and 3 that the physical curve is the one on
the right side of the right singularity, i.e., the physical region
is bounded by a temperature, which varies according to the
constant over the fine structure constant under the logarithm.
With detailed observation on the physical curve, one finds that it
does not decrease monotonically with the temperature but presents
a minimum value of about 0.4 at $T=182.1MeV$ instead, which is
twice greater than the phenomenological estimation\cite{Teaney}.
This minimum value, seeing from mathematical point of view, is
owing to the competition between the weakening running coupling
constant and the rising temperature itself.

\section{Summary}
In the framework of irreversible thermodynamics, the transport
properties of so-called QGP produced in heavy ion collision have
been studied. Based on the Boltzmann equation, we calculated the
shear viscosity of two-flavor QGP at high temperature and finite
density in weakly coupled limit in kinetic theory. The result
shows the finite density effect provides positive contribution to
the shear viscosity by adding a quadratic term of $\mu/T$ to the
pure temperature case. Furthermore, we calculated the
non-equilibrium entropy density at finite chemical potential
through two different schemes by subtracting the entropy
production due to shear viscous process from the final equilibrium
entropy density. In our evaluation, the thermal force in the shear
viscous process, namely, the velocity gradient, was estimated by
it maximum value considering the longitudinal expansion in heavy
ion collision. Finally the viscosity to entropy density ratio was
demonstrated in a extensive temperature regime including both
physical and unphysical regions, which are identified by two
natural conditions, to see clearly the temperature boundary. In
the physical region the ratio appears a minimum value of 0.4 which
is two times larger than that expected from the ideal behavior of
the elliptic flow. The chemical potential effect, distinguished
from the viscosity case, decreases the entropy density and thus
contributes a final positive effect on the ratio.

At last, we emphasize here that one should pay attention not only
to the ratio of transport coefficients to the equilibrium entropy
density but also to the ratio of transport coefficients to the
non-equilibrium entropy density when to understand the transport
or dissipative properties of QGP. Generally speaking, the entropy
production $dS$ comes either from the external sources $d_{ex}S$
and the internal dissipative and viscous processes, which is
expressed as
\begin{equation}
dS=d_{in}S +d_{ex}S.
\end{equation}
As to the external sources, two basic aspects should be taken into
account. One is the energy loss of parton when it is passing
through the medium. The other might be the increase of the degree
of freedom excited by the phase transition. Actually, these
problems are really open and far from known, which
need further study.\\[0.5cm]

\centerline{\bf Acknowledgement} This work is partly supported by
the National Natural Science Foundation of China under project
Nos. 90303007, 10135030 and 10575043, the Ministry of Education of
China with Project No. CFKSTIP-704035.

\
 \\[1cm]
\begin{appendix}
\noindent {\large\bf Appendix: Relaxation time}
\\[1cm]
In this appendix, we first define the relaxation time in Boltzmann
equation, and then relate it with the shear viscosity which has
been already obtained in the variational approach.

The Boltzmann equation without time derivative and external force
terms in the relaxation time approximation is to substitute the
collision term on the right hand side for the fluctuation of the
distribution function scaled by the relaxation time $\tau_{\eta}$
\begin{equation}
\mathbf{\hat p}\cdot \frac{\partial f_s^0(\mathbf{x;p})}{\partial
\mathbf{x}} =-\frac{\delta f_s}{\tau_{\eta}},
\end{equation}
where $f^0(\mathbf{x;p})=[\exp (-\beta P_\nu u^\nu \mp \mu)\pm
1]^2$ is the J\"{u}ttner distribution function\cite{Groot}. With
this definition and Eq. (\ref{viscosity definition}), one could
find
\begin{eqnarray}
\delta \langle
T_{zx}\rangle=&&\hspace{-0.6cm}\tau_{\eta}\beta\int_{\mathbf{p}}
\left(\frac{p_x
p_z}{p^2}\right)^2 \big\{g_f n_f(p)[1-n_f(p)] \nonumber\\
&&+g_{\bar f} n_{\bar f}(p)[1-n_{\bar f}(p)]+g_b
n_b(p)[1+n_b(p)]\big\}\left(\frac{\partial u_z}{\partial
x}\right)\\
&&\hspace{-0.6cm}=\eta \frac{\partial u_z}{\partial x}
\end{eqnarray}
where the last line is the linear law.

Noticing the velocity gradient is independent with momentum thus
can be cancelled on both side, we can find out the relation
between the relaxation time and the shear viscosity by trivially
evaluating the momentum integral,
\begin{equation}
\tau_{\eta}=\eta\left[3.246T^4\left(1+0.246\frac{\mu^2}{T^2}\right)\right]^{-1}.
\end{equation}

\end{appendix}

\end{document}